\documentclass[11pt,a4paper]{article}

\usepackage[utf8]{inputenc}
\usepackage[margin=1in]{geometry}
\usepackage{amsmath,amssymb,amsfonts,bm}
\usepackage{graphicx}
\usepackage{booktabs}
\usepackage{tabularx}
\usepackage{array}
\usepackage[section]{placeins}
\usepackage[square,numbers,sort&compress]{natbib}
\usepackage{microtype}
\usepackage{xcolor}
\usepackage{tikz}
\usetikzlibrary{shapes.geometric, positioning, arrows.meta, calc}
\usepackage[colorlinks=true,linkcolor=blue,citecolor=blue,urlcolor=blue]{hyperref}

\title{\Large \textbf{AI Meets Turbulence Modeling: Balancing Causation, Correlation, and User Intent}}

\author{
  \textbf{Sharath S. Girimaji}\thanks{Corresponding author: \href{mailto:girimaji@tamu.edu}{girimaji@tamu.edu}}\\[0.5em]
  \small Department of Ocean Engineering, Department of Aerospace Engineering,\\
  \small Texas A\&M University, College Station, TX 77843, USA
}

\date{\today}

\begin{document}

\maketitle

\begin{abstract}
Turbulence states vary dramatically in modeling difficulty, mandating fundamentally different modeling paradigms; consequently, it is increasingly evident that no single artificial intelligence architecture or data-driven approach is universally suited for all flows of interest. This manuscript proposes a comprehensive operational framework that classifies data-driven turbulence modeling strategies by balancing physical causation and statistical correlation according to non-equilibrium flow memory along with user intent. Using the Mori--Zwanzig projection formalism, we decompose the unclosed turbulent stress into a local response ($\tau_{ij}^L$) and a spatio-temporally non-local memory integral ($\tau_{ij}^{NL}$). In flows near statistical equilibrium, rapid flow state relaxation suppresses advective memory, allowing local statistical correlation to capture physical causation via constitutive relations for $\tau_{ij}^L$ implemented intrusively within Reynolds-Averaged Navier--Stokes (RANS) equations. Crucially, the elliptic Poisson equation for fluctuating pressure couples stress evolution to eddy dimensionality beyond single-point velocity-gradient componentiality. Therefore, local constitutive closures are not strictly valid even in equilibrium turbulence, yet they serve as an acceptable engineering approximation. In non-equilibrium turbulence, the modeling strategy bifurcates according to user intent. For predictive calculations of unseen geometries, large coherent structures sustain non-local memory ($\tau_{ij}^{NL}$), invalidating local correlation-based closures for the total stress and mandating first-principles causal scale resolution guided by Rules-Augmented AI (RAAI). Resolving these macroscopic structures directly captures non-local dynamics, with a recently proposed non-equilibrium theory determining the precise degree of resolution required. Conversely, for rapid design within a known, parameterized coherent flow category, a purely correlational, data-driven strategy suffices. Combining Non-Intrusive Reduced-Order Modeling (NIROM) with uncentered Proper Orthogonal Decomposition (POD) across multi-realization data ensembles enables zero-solver synthesis of both stationary statistics and unsteady coherent time evolution directly from governing dimensionless parameters. Ultimately, this multi-pronged strategy offers a balanced path forward by aligning physical causation, statistical correlation, and user intent across diverse flow regimes.
\end{abstract}

\section{Introduction}
\label{sec:intro}

Many engineering and geophysical problems entail predicting unseen turbulent flows dominated by energetic coherent structures that challenge conventional closure models. In contrast, numerous practical design and optimization workflows do not need full unsteady simulations across novel geometries; instead, they require characterization of statistically stationary mean velocity fields, Reynolds stress tensors, or coherent unsteady dynamics within a known, parameterized flow family. Addressing these distinct goals across varying flow complexities requires replacing ad-hoc surrogate models with a physics-based taxonomy that matches modeling fidelity to the minimum physical requirements of the task \cite{girimaji2026nsr}. Purely correlation-driven models can suffice in certain regimes, whereas causal formulations are required in others. When closures are deployed on unseen flows, generalization depends on the ability of the data-driven models to capture physical causation through correlation. Effective data-driven modeling therefore hinges on the underlying flow physics: distinct dynamical regimes present fundamentally different modeling challenges that no single machine learning architecture can optimally handle, making a strategy that balances physical causation, statistical correlation, and user intent essential \cite{girimaji2024njp}.

The physical admissibility of local closures, and hence the ability of statistical correlations to capture causality, depends directly on how close the flow is to the equilibrium state \cite{girimaji2024njp}. Here, equilibrium denotes a balance between linear mean-flow forcing (production) and non-linear relaxation (dissipation), characterized by the turbulent Deborah number, $De_{\text{turb}} \equiv S k / \epsilon$. This parameter compares the non-linear eddy relaxation timescale ($\tau_N = k/\epsilon$) to the characteristic mean strain timescale ($\tau_L = 1/S$) and scales monotonically with the production-to-dissipation ratio \cite{monin1971, girimaji2026nsr, girimaji2026neq}. In near-equilibrium turbulence (Regimes~I and~II of \cite{girimaji2026nsr}), non-linear eddy interactions relax faster than the mean deformation evolves, attenuating advective history. 
Under these conditions, local production--dissipation balance ties causal stress dynamics to local statistical correlations, making a local constitutive approximation reasonable \cite{girimaji2024njp}. Even near equilibrium, however, the eddy-dimensionality tensor can significantly influence stress evolution, requiring the constitutive relation to incorporate structural information beyond the mean velocity gradients ($\nabla \bar{\mathbf{u}}$) \cite{kassinos1994, kassinos2001}.

Flows far from equilibrium (Regimes~III and~IV) exhibit strong spatio-temporal memory---arising from coarse-graining over energetic coherent structures---that extends beyond elliptic pressure (dimensionality tensor) effects. This persistent non-Markovian history breaks standard local correlation-based closures for the total turbulent stress, preventing them from inferring physical causation from local flow statistics alone. Moreover, because these memory effects depend on the specific coherent structures present, data-driven closures trained on one flow category generally do not transfer to another \cite{girimaji2026nsr}.

Navigating this non-equilibrium regime requires adapting the modeling strategy according to user intent---separating predictive extrapolation to novel geometries from within-class design interpolation \cite{girimaji2026nsr}. For predictive simulations of unseen geometries, non-local memory effects cannot be inferred from prior data, mandating first-principles causal scale resolution guided by Rules-Augmented AI (RAAI). A recent non-equilibrium turbulence theory \cite{girimaji2026neq} establishes the precise degree of scale resolution required on the basis of the unresolved production-to-dissipation ratio ($P_u/\epsilon_u$, directly tied to the subgrid Deborah number) as the governing thermodynamic order parameter. On the other hand, for repetitive design simulations within a known, parameterized flow category $\mathcal{T}$, a purely correlational, data-driven strategy suffices: non-intrusive reduced-order modeling (NIROM) combined with collective, uncentered Proper Orthogonal Decomposition (POD) across multi-realization data ensembles. This approach deliberately trades physical causality for computational speed, enabling zero-solver characterization of stationary flow statistics and unsteady coherent-structure time evolution directly as a function of governing dimensionless parameters \cite{hesthaven2018, girimaji2026nsr}. However, realizing this within-class efficiency without risking unphysical extrapolation requires establishing mathematically rigorous criteria to define a valid flow category.

In this manuscript, we organize data-driven turbulence modeling into a balanced, physics-governed framework. Section~\ref{sec:mz} develops the Mori--Zwanzig projection-operator decomposition into local ($\tau_{ij}^L$) and spatio-temporally non-local ($\tau_{ij}^{NL}$) stress contributions, establishing the statistical-mechanical foundations of local constitutive modeling and separating intrusive versus non-intrusive architectures. Section~\ref{sec:equilibrium} details the modeling requirements for near-equilibrium regimes, covering the tensor integrity-basis closure for total Reynolds stresses---explicitly addressing its physical incompleteness regarding eddy dimensionality arising from the elliptic pressure Poisson equation \cite{kassinos1994, kassinos2001}. Section~\ref{sec:nonequilibrium} addresses non-equilibrium regimes, showing how RAAI-guided scale resolution in unseen geometries captures non-Markovian memory so the classic tensor expansion serves as a causal subgrid closure. This section also details the NIROM recipe for stationary synthesis and unsteady animation in seen flow families. Importantly, this section establishes the four formal mathematical criteria defining a valid Flow Category $\mathcal{T}$. Finally, Section~\ref{sec:roadmap} synthesizes these results into an operational decision roadmap.

\section{Mori--Zwanzig Framework: Local Response versus Spatio-Temporal Non-Locality}
\label{sec:mz}


To establish a rigorous mathematical boundary between local correlation and non-local causation, we examine the unclosed turbulent stress tensor $\tau_{ij}(\mathbf{x},t)$ through the Mori--Zwanzig (MZ) projection-operator formalism \cite{zwanzig1961, zwanzig2001, chorin2002, parish2017}. Let $\mathbf{u}(\mathbf{x},t)$ denote the full turbulent velocity field governed by the Navier--Stokes evolution operator $\mathcal{L}$, partitioned via a projection operator $\mathcal{P}$ onto the resolved subspace and its orthogonal complement $\mathcal{Q}$ onto the unresolved fluctuations:
\begin{equation}
    \frac{\partial \mathbf{u}}{\partial t} = \mathcal{L}\mathbf{u}, \qquad \bar{\mathbf{u}} = \mathcal{P}\mathbf{u}, \qquad \mathbf{u}' = \mathcal{Q}\mathbf{u} = (\mathcal{I} - \mathcal{P})\mathbf{u}.
    \label{eq:mz_projection}
\end{equation}
Here, $\mathcal{P}$ represents a general scale-separation operator that may correspond either to a full Reynolds ensemble average (RANS) or to a spatial/spectral scale-resolving filter (such as PANS or LES). Because the full evolution operator splits into resolved and unresolved pathways ($\mathcal{L} = \mathcal{P}\mathcal{L} + \mathcal{Q}\mathcal{L}$), the full propagator $e^{t\mathcal{L}}$ expands exactly via Dyson's operator identity (the operator equivalent of Duhamel's principle) \cite{zwanzig2001, parish2017}:
\begin{equation}
    e^{t\mathcal{L}} = e^{t\mathcal{Q}\mathcal{L}} + \int_0^t e^{(t-s)\mathcal{L}}\, \mathcal{P}\mathcal{L}\, e^{s\mathcal{Q}\mathcal{L}}\, ds.
    \label{eq:dyson}
\end{equation}
Applying Eq.~\eqref{eq:dyson} to the unclosed nonlinear interactions (and omitting the purely orthogonal initial-condition transient) partitions the deterministic turbulent stress tensor into two distinct physical contributions:
\begin{equation}
    \tau_{ij}(\mathbf{x},t) = \tau_{ij}^L(\mathbf{x},t) + \tau_{ij}^{NL}(\mathbf{x},t),
    \label{eq:mz_stress}
\end{equation}
where $\tau_{ij}^L(\mathbf{x},t)$ denotes the \emph{local response} and $\tau_{ij}^{NL}(\mathbf{x},t)$ represents the \emph{spatio-temporally non-local memory integral}.

The local component $\tau_{ij}^L(\mathbf{x},t)$ arises from the immediate projection $\mathcal{P}\mathcal{L}$ of nonlinear interactions directly onto the resolved manifold at $(\mathbf{x},t)$, bypassing any dynamical excursion through the orthogonal subspace $\mathcal{Q}$. Because this term adjusts instantaneously to the local resolved state without retaining advective history, it admits a pointwise algebraic constitutive representation:
\begin{equation}
    \tau_{ij}^L(\mathbf{x},t) = \mathcal{M}_{ij}\!\left(\nabla \bar{\mathbf{u}}(\mathbf{x},t),\, k(\mathbf{x},t),\, \epsilon(\mathbf{x},t)\right),
    \label{eq:local_stress}
\end{equation}
where $\mathcal{M}_{ij}$ is an isotropic tensor-valued constitutive operator parameterized exclusively by the local resolved velocity gradient $\nabla \bar{\mathbf{u}}(\mathbf{x},t)$ and local turbulence scales ($k, \epsilon$). Conversely, the non-local component $\tau_{ij}^{NL}(\mathbf{x},t)$ originates from the integral term in Eq.~\eqref{eq:dyson}, which tracks a delayed round-trip interaction: resolved-scale straining at past times $s < t$ and remote locations $\mathbf{x}' \in \Omega$ excites unresolved fluctuations that propagate along the orthogonal propagator $\exp[(t-s)\mathcal{Q}\mathcal{L}]$---undergoing Lagrangian advection, vortex stretching, elliptic pressure coupling, and boundary reflections---before feeding back onto the resolved stress at $(\mathbf{x},t)$. Integrating this dynamical history over space and past time yields the explicit convolution representation:
\begin{equation}
    \tau_{ij}^{NL}(\mathbf{x},t) = \int_0^t \int_{\Omega} K_{ij}[\mathcal{F};\, \mathbf{x}, \mathbf{x}', t, s]\, \bar{\mathbf{u}}(\mathbf{x}', s)\, d\mathbf{x}' \, ds,
    \label{eq:mz_kernel}
\end{equation}
where the non-local memory kernel $K_{ij}[\mathcal{F};\, \mathbf{x}, \mathbf{x}', t, s]$ represents the Green's-function response of the orthogonal propagator $\exp[(t-s)\mathcal{Q}\mathcal{L}]$. Because orthogonal trajectories are constrained by the global domain geometry, boundary conditions, and large-scale linear instabilities encapsulated by $\mathcal{F}$ \cite{chorin2002, parish2017}, the functional form of $K_{ij}$ is inherently global and flow-specific.

The relative dominance of $\tau_{ij}^L$ versus $\tau_{ij}^{NL}$ is governed by the turbulent Deborah number of the unresolved scales in $\mathcal{Q}$, comparing the non-linear eddy relaxation timescale ($\tau_N = k/\epsilon$) to the characteristic mean deformation timescale ($\tau_L = \mathcal{O}(1/S)$):
\begin{equation}
    De_{\text{turb}} \equiv \frac{S k}{\epsilon} = \frac{\tau_N}{\tau_L}.
    \label{eq:de_turb}
\end{equation}
When $De_{\text{turb}} \ll 1$, rapid nonlinear eddy relaxation decorrelates the orthogonal propagator $\exp[(t-s)\mathcal{Q}\mathcal{L}]$ on a timescale much shorter than mean flow evolution, causing advective memory effects to decay rapidly and collapsing Eq.~\eqref{eq:mz_stress} onto its local leading order, $\tau_{ij}(\mathbf{x},t) \approx \tau_{ij}^L(\mathbf{x},t)$ \cite{pope2000}. Three foundational pillars of non-equilibrium statistical mechanics explain why $\tau_{ij}^L$ admits a local algebraic constitutive closure in this limit, whereas $\tau_{ij}^{NL}$ cannot:
\begin{enumerate}
    \item \textbf{Gas Kinetic Theory and Scale Separation:} Just as molecular collisions in the Boltzmann equation relax the velocity distribution toward a local Maxwellian over a mean free path much shorter than macroscopic flow scales (the small Knudsen number limit, $Kn \ll 1$), rapid nonlinear triad interactions in near-equilibrium turbulence ($Kn_{\text{turb}} \sim De_{\text{turb}} \ll 1$) equilibrate unresolved fluctuations locally before fluid elements traverse macroscopic geometric gradients \cite{monin1971, pope2000, bgk1954}. In single-point Reynolds-averaged form, this scale separation corresponds to Rodi's weak-equilibrium balance ($P_{ij} + \Pi_{ij} - \epsilon_{ij} = \frac{\tau_{ij}}{k}(P-\epsilon)$), which underpins Algebraic Reynolds Stress Models \cite{rodi1976, girimaji1996}.
    \item \textbf{Chapman--Enskog Asymptotic Analysis:} Expanding the unresolved subgrid dynamics in powers of the small scale-separation parameter $\varepsilon \sim De_{\text{turb}}$ systematically eliminates material time-history derivatives at successive asymptotic orders \cite{chapman1970, girimaji2024njp}. At zeroth order ($\mathcal{O}(1)$), the unresolved stress is locally isotropic; at first ($\mathcal{O}(\varepsilon)$) and higher algebraic orders in the Chapman--Enskog series, advective memory dependencies collapse onto local spatial gradients of the resolved velocity field. This asymptotic reduction shows that $\tau_{ij}^L$ is governed predominantly by instantaneous local velocity gradients $\nabla\bar{\mathbf{u}}(\mathbf{x},t)$ and local scalar scales ($k,\epsilon$), allowing statistical correlation to absorb local physical causation.
    \item \textbf{Onsager's Near-Equilibrium Thermodynamics and Reciprocity:} Around a state of local statistical equilibrium, Onsager's regression hypothesis and reciprocal relations dictate that macroscopic fluxes (turbulent momentum fluxes $\tau_{ij}^L$) respond locally to their conjugate thermodynamic forces (mean strain-rate gradients $\nabla\bar{\mathbf{u}}$) via symmetric, positive-definite transport coefficients governed by local state variables \cite{onsager1931}. Because local dissipative relaxation dominates advective transport, these constitutive coefficients remain largely independent of global boundary conditions $\mathcal{F}$ and generalize across geometries.
\end{enumerate}

Even in the near-equilibrium regime ($De_{\text{turb}} \ll 1$), however, incompressible turbulence retains an intrinsic source of spatial non-locality through the fluctuating pressure field, which is governed by an elliptic Poisson equation rather than a local hyperbolic relaxation law. Consequently, while $\tau_{ij}^L$ provides an indispensable engineering closure, no single-point algebraic function $\mathcal{M}_{ij}(\nabla\bar{\mathbf{u}}, k, \epsilon)$ can represent equilibrium stresses with complete mathematical exactness---a fundamental structural limitation examined in Section~\ref{sec:equilibrium}. Far from equilibrium ($De_{\text{turb}} \gg 1$), linear mean deformation outpaces nonlinear relaxation ($\tau_N \gg \tau_L$): the Chapman--Enskog series diverges, Onsager's local flux--force reciprocity breaks down, and substantial-derivative advection ($D\tau_{ij}/Dt$), spatial transport, and elliptic pressure--Poisson coupling sustain long-range kinematic history across the entire domain $\Omega$ \cite{girimaji2024njp, girimaji2026nsr}. Because $\tau_{ij}^{NL}$ in Eq.~\eqref{eq:mz_kernel} is a global spatio-temporal functional of $\bar{\mathbf{u}}(\mathbf{x}', s)$ mediated by a geometry-dependent kernel ($K_{ij}[\mathcal{F}_1] \neq K_{ij}[\mathcal{F}_2]$) \cite{zwanzig1961, gouasmi2017}, no pointwise algebraic constitutive relation of the form Eq.~\eqref{eq:local_stress} exists for $\tau_{ij}^{NL}$. Data-driven closures conditioned solely on local velocity gradients or invariants remain blind to $\mathcal{F}$ and attempt a non-injective mapping---forcing distinct non-local flow histories onto identical local inputs and incurring an irreducible Bayes error \cite{taghizadeh2021pof, girimaji2024njp}. Resolving this impasse requires either redefining $\mathcal{P}$ to resolve the memory-carrying structures explicitly so that the residual subgrid stress reverts to $\tau_{ij}^L$, or confining purely correlational surrogates to an invariant flow category.

\subsection{Intrusive versus Non-Intrusive Modeling Paradigms}

Interpreting the MZ decomposition operationally reveals two parallel modeling pathways: a \emph{memory-driven, causal, intrusive} pathway and a \emph{dynamically memoryless, correlational, non-intrusive} pathway. When non-local memory $\tau_{ij}^{NL}$ is active and unknown (unseen flows at $De_{\text{turb}} \gg 1$), reconstructing the spatio-temporal history in Eq.~\eqref{eq:mz_kernel} requires preserving physical causation by marching the governing partial differential equations forward in time, making any embedded closure or scale-resolution arbitrator inherently \emph{intrusive}. Conversely, when restricted to a known Flow Category $\mathcal{T}$ across which $K_{ij}[\mathcal{F}]$ remains topodynamically invariant, the surrogate bypasses temporal trajectory integration altogether---yielding a dynamically \emph{memoryless}, purely correlational parameter-to-field mapping that is entirely \emph{non-intrusive}.

This structural bifurcation dictates fundamentally different mathematical requirements for \emph{intrusive} versus \emph{non-intrusive} models (summarized in Table~\ref{tab:intrusive_vs_nonintrusive}). Intrusive closures are embedded directly within the governing partial differential equations, operating alongside convective, diffusive, and pressure-gradient operators to evolve the flow field in time or iterate toward statistical stationarity. Because an intrusive model actively governs how the flow state advances from one instant to the next, it carries the responsibility of representing physical causation within the coupled system. Consequently, an intrusive closure must possess both \emph{kinematic} and \emph{dynamic} characteristics strictly consistent with the underlying conservation laws. Kinematically, the modeled Reynolds-stress tensor must remain symmetric and positive semi-definite (realizable) while adhering to domain boundary conditions. Dynamically, the closure must obey Galilean invariance, rotational covariance, appropriate asymptotic limits (such as Rapid Distortion Theory and two-component near-wall limits), and compatibility with exact energy and stress transport balances. If these dynamic properties are violated, a data-driven model that yields reasonable predictions at an isolated time instant (\emph{a priori} accuracy) can trigger spurious nonlinear feedback loops, numerical stiffness, or severe trajectory drift when coupled into the differential equation system to evolve the flow field (\emph{a posteriori} fidelity).

In contrast, non-intrusive models---such as Proper Orthogonal Decomposition (POD) surrogates and parameter-to-field regressions---function entirely outside the differential solver, mapping non-dimensional geometric and operating parameters directly to statistically stationary flow fields and attractor trajectories. Because non-intrusive surrogates perform no temporal PDE integration and do not interact with differential transport operators, they carry no causal burden and are exempt from dynamic evolution constraints. Instead, their admissibility is governed strictly by \emph{kinematic} constraints: the reconstructed velocity field must conserve mass (remaining divergence-free in incompressible flows) and satisfy geometric boundary conditions, while the synthesized Reynolds-stress tensor must preserve symmetry and realizability across the parameter domain. Separating the joint kinematic--dynamic requirements of intrusive closures from the purely kinematic requirements of non-intrusive surrogates establishes minimal, sufficient design criteria for model construction across both paradigms.

\begin{table}[htbp]
\centering
\caption{Operational and mathematical comparison of intrusive closures versus non-intrusive surrogates.}
\label{tab:intrusive_vs_nonintrusive}
\footnotesize
\renewcommand{\arraystretch}{1.35}
\setlength{\tabcolsep}{6pt}
\begin{tabularx}{\textwidth}{>{\bfseries\raggedright\arraybackslash}p{0.19\textwidth} >{\raggedright\arraybackslash}X >{\raggedright\arraybackslash}X}
\toprule
\textbf{Attribute} & \textbf{Intrusive Models} & \textbf{Non-Intrusive Models} \\
\midrule
Conceptual Pathway & Memory-driven $\rightarrow$ Causal $\rightarrow$ Intrusive & Dynamically memoryless $\rightarrow$ Correlational $\rightarrow$ Non-intrusive \\
Solver Coupling & Embedded directly inside governing PDEs alongside convective, diffusive, and pressure operators. & External to PDE solver; direct algebraic mapping from non-dimensional parameters $\boldsymbol{\Pi}$ to stationary fields and attractor modes. \\
Causal Responsibility & High: governs step-by-step state advancement and trajectory evolution (\emph{a posteriori} fidelity). & None: zero physical causality; interpolates pre-computed manifold solutions within Flow Category $\mathcal{T}$. \\
Dynamic Constraints & \textbf{Mandatory:} Galilean invariance, rotational covariance, material frame indifference, RDT/2C asymptotic limits, and energy/transport balance compatibility. & \textbf{Exempt:} No differential transport laws or PDE operator coupling required. \\
Kinematic Constraints & \textbf{Mandatory:} Stress tensor symmetry, positive semi-definiteness (realizability), and boundary conditions. & \textbf{Mandatory:} Mass conservation ($\nabla \cdot \bar{\mathbf{u}} = 0$), boundary conditions, and Reynolds-stress symmetry/realizability. \\
Primary Failure Mode & Spurious nonlinear feedback, numerical stiffness, or trajectory divergence during coupled solver integration. & Unphysical extrapolation when applied outside the trained Flow Category $\mathcal{T}$ or across bifurcations. \\
\bottomrule
\end{tabularx}
\end{table}

\section{Near-Equilibrium Turbulence ($De_{\text{turb}} \ll 1$): Intrusive Correlation Models}
\label{sec:equilibrium}

\subsection{RDT Sampling Bounds and State-Space Coverage}

In the near-equilibrium regime ($De_{\text{turb}} \ll 1$; Regimes~I and~II), $\mathcal{P}$ represents the full Reynolds ensemble average and rapid nonlinear eddy relaxation justifies approximating the \emph{total} Reynolds stress via its local constitutive leading order, $\tau_{ij} \approx \tau_{ij}^L$. Within this regime, classical algebraic closures act as the turbulence analog of the Bhatnagar--Gross--Krook (BGK) single-relaxation-time approximation in kinetic theory \cite{bgk1954}, relaxing anisotropic stresses toward equilibrium at the local eddy turnover frequency $\epsilon/k$. Because local statistical correlation absorbs local physical causation when advective memory vanishes, data-driven constitutive models trained on high-fidelity reference data can generalize across unseen geometries---provided the training dataset spans the admissible state space without bias.

Rapid Distortion Theory (RDT) provides the rigorous mathematical foundation for establishing these training bounds \cite{batchelor1954, townsend1976, kassinos2001}. By linearizing the fluctuating Navier--Stokes equations about mean strain-rate ($S_{ij}$) and rotation-rate ($\Omega_{ij}$) tensors, RDT isolates the fundamental linear mechanisms of vortex stretching and rapid pressure--strain redistribution. Analyzing the asymptotic fixed points and bifurcation boundaries of RDT defines the exact multi-dimensional parameter bounds---expressed in terms of strain-to-rotation invariant ratios ($II_S / II_\Omega$) and normalized Reynolds-stress anisotropy states ($b_{ij}$)---required to systematically sample data-driven closures while avoiding redundant over-sampling of saturated linear regimes \cite{girimaji2026nsr}.

\subsection{Ten-Tensor Integrity-Basis Closure: Componential Approximation and Intrusive Constraints}

In Regimes~I and~II, the suppression of advective memory allows the \emph{total} Reynolds-stress anisotropy tensor $b_{ij} \equiv \langle u_i' u_j' \rangle / (2k) - \frac{1}{3}\delta_{ij}$ to be modeled causally via Pope's complete ten-tensor integrity-basis expansion \cite{pope1975, taghizadeh2020njp}:
\begin{equation}
    b_{ij} = \sum_{n=1}^{10} G_n(\mu_1, \dots, \mu_5)\, T_{ij}^{(n)}(\hat{\mathbf{S}}, \hat{\boldsymbol{\Omega}}),
    \label{eq:rans_10tensor}
\end{equation}
where $T_{ij}^{(n)}$ ($n = 1, \dots, 10$) form the complete, linearly independent symmetric deviatoric tensor basis constructed from normalized mean strain-rate ($\hat{\mathbf{S}}$) and rotation-rate ($\hat{\boldsymbol{\Omega}}$) tensors, and $G_n$ are scalar basis functions of the five independent local invariants ($\mu_1, \dots, \mu_5$) learned from RDT-bounded training data.

Crucially, however, fundamental turbulence structure theory demonstrates that Eq.~\eqref{eq:rans_10tensor} is \textbf{not a complete physical representation of turbulent stress, even in strictly equilibrium turbulence} \cite{kassinos2001, mishra2010, mishra2013jfm, mishra2017jfm}. Because the ten integrity-basis tensors $T_{ij}^{(n)}$ are constructed exclusively from single-point mean velocity gradients ($\hat{\mathbf{S}}, \hat{\boldsymbol{\Omega}}$), they parameterize only \textbf{componentiality}---the directional partitioning of turbulent kinetic energy among velocity components ($\langle u_i' u_j' \rangle$). As established by Kassinos, Reynolds, and Rogers \cite{kassinos2001}, a complete one-point description of turbulence dynamics also depends on eddy \textbf{dimensionality} ($D_{ij}$, alongside circulicity $F_{ij}$), which characterizes the spatial morphology, aspect ratio, and spectral wavevector alignment of the turbulent eddies.

This dependence on dimensionality arises directly from the \textbf{elliptic Poisson equation} governing fluctuating pressure in incompressible flows:
\begin{equation}
    \nabla^2 p' = -2\rho\, \frac{\partial \bar{u}_i}{\partial x_j} \frac{\partial u_j'}{\partial x_i} - \rho\, \frac{\partial^2}{\partial x_i \partial x_j}\!\left(u_i' u_j' - \langle u_i' u_j' \rangle\right).
    \label{eq:pressure_poisson}
\end{equation}
Because the Green's-function solution of Eq.~\eqref{eq:pressure_poisson} integrates velocity fluctuations globally over physical space---or equivalently couples wavevector orientations in spectral space---inter-component energy transfer mediated by the pressure--strain correlation depends sensitively on eddy dimensionality \cite{mishra2010, mishra2013jfm}. As demonstrated by Mishra and Girimaji \cite{mishra2013jfm, mishra2017jfm}, turbulent flows sharing identical local velocity gradients ($\hat{\mathbf{S}}, \hat{\boldsymbol{\Omega}}$) and even identical componentiality $b_{ij}$ undergo measurably different pressure--strain redistributions if their underlying dimensionality tensors differ. Because single-point velocity gradients carry no direct information about eddy dimensionality, the ten-tensor expansion in Eq.~\eqref{eq:rans_10tensor} cannot capture equilibrium Reynolds stresses with exact precision; rather, in near-equilibrium turbulence ($De_{\text{turb}} \ll 1$), where rapid nonlinear interactions scramble wavevector orientations enough that dimensionality tracks local strain history closely, Eq.~\eqref{eq:rans_10tensor} serves as an \textbf{acceptable engineering approximation} subject to an irreducible single-point epistemic floor.

When Eq.~\eqref{eq:rans_10tensor} is deployed \emph{intrusively} inside the Reynolds-averaged momentum equations to update velocity and turbulence fields dynamically, \emph{a priori} regression accuracy on static snapshots is insufficient to guarantee stable \emph{a posteriori} predictions. Intrusive integration strictly requires embedding three physical dynamic and kinematic constraints directly into the model architecture:
\begin{enumerate}
    \item \textbf{Galilean Invariance:} Constitutive relations must depend exclusively on spatial velocity gradients and local scalar invariants rather than raw velocity vectors, ensuring strict invariance under constant-velocity frame translations \cite{speziale1987}.
    \item \textbf{Material Frame Indifference and Rotational Covariance:} The ten-tensor integrity basis guarantees objective transformation under three-dimensional coordinate rotations, while the learned scalar coefficients $G_n$ must additionally respect two-dimensional material frame indifference in the two-component near-wall asymptotic limit \cite{speziale1987, taghizadeh2020njp}.
    \item \textbf{Comprehensive Realizability:} Predicted Reynolds-stress tensors must remain symmetric and positive semi-definite ($\tau_{ij} \in \mathcal{S}_+^3$, with non-negative eigenvalues and bounded anisotropy invariants within the Lumley triangle) at every spatial point and integration step to prevent unphysical negative normal stresses and numerical divergence during closed-loop PDE solution \cite{lumley1978, girimaji2004jfm}.
\end{enumerate}

\section{Non-Equilibrium Turbulence ($De_{\text{turb}} \gg 1$): Unseen Topologies vs. Seen Flow Categories}
\label{sec:nonequilibrium}

Far from statistical equilibrium ($De_{\text{turb}} \gg 1$; Regimes~III and~IV), mean deformation outpaces nonlinear eddy relaxation, causing the spatio-temporally non-local Mori--Zwanzig memory integral $\tau_{ij}^{NL}$ in Eq.~\eqref{eq:mz_kernel} to dominate stress transport. Under a full RANS ensemble average, large memory-carrying coherent structures reside inside the unresolved orthogonal subspace $\mathcal{Q}$, invalidating the local ten-tensor algebraic expansion in Eq.~\eqref{eq:rans_10tensor} for the \emph{total} turbulent stress. Navigating this non-equilibrium regime therefore bifurcates into two distinct pathways governed by computational intent: causal scale resolution for unseen topologies versus topological memorization within a certified flow category.

\subsection{Unseen Topologies: Causal Scale Resolution and the Ten-Tensor Subgrid Closure}

For predictive calculations of \textbf{unseen flow geometries} ($\mathcal{F}_{\text{unseen}}$), the global memory kernel $K_{ij}[\mathcal{F}_{\text{unseen}}]$ is physically unknown and cannot be inferred from prior training datasets. Preserving physical causation therefore mandates \emph{Pathway~(i)}: first-principles, causal scale-resolving simulations---such as Partially-Averaged Navier--Stokes (PANS) or Large-Eddy Simulation (LES)---that redefine the projection operator $\mathcal{P}$ as a spatial/spectral filter and directly integrate the unsteady Navier--Stokes equations to resolve the energy-containing, memory-bearing coherent structures \cite{sagaut2006, girimaji2006pans}. By transferring the non-local trajectory dynamics of macroscopic vortices from $\mathcal{Q}$ into the resolved subspace $\mathcal{P}$, causal scale resolution shrinks the unresolved subgrid subspace until the residual small scales revert to local statistical equilibrium ($De_{\text{turb}}|_{\text{subgrid}} \ll 1$) \cite{taghizadeh2024pof}.

Deploying scale resolution efficiently without resorting to prohibitively expensive Direct Numerical Simulation requires adaptive, \emph{in situ} filter control. In a Rules-Augmented AI (RAAI) bridging framework, an embedded machine-learning agent acts as a local resolution arbitrator, dynamically modulating the unresolved-to-total kinetic energy and dissipation ratios ($f_k, f_\epsilon$) in response to local non-equilibrium indicators such as $P/\epsilon$ and $De_{\text{turb}}$ \cite{girimaji2006pans, taghizadeh2024pof}. Where coherence and non-equilibrium memory are intense, the arbitrator depresses $f_k \to 0$ to resolve macroscopic structures explicitly; in quiescent or near-equilibrium regions, it relaxes $f_k \to 1$ to recover economical RANS efficiency.

Crucially, explicitly resolving the energy-containing eddies inside $\mathcal{P}$ accomplishes two things simultaneously: it eliminates advective memory from the unresolved subspace ($\tau_{ij}^{NL}|_{\text{subgrid}} \to 0$) and \textbf{directly resolves the anisotropic eddy dimensionality and global elliptic pressure field of the large structures on the computational grid}. Consequently, the residual subfilter eddies left in $\mathcal{Q}$ are stripped of large-scale directional morphology and approach small-scale isotropy, where \textbf{the exact same ten-tensor integrity-basis expansion} from Section~\ref{sec:equilibrium}---now generalized with cut-off-dependent scalar basis functions $G_n(\mu_1, \dots, \mu_5, f_k)$---becomes an accurate \emph{causal subfilter closure} for the unresolved stress anisotropy \cite{girimaji2006pans, taghizadeh2024pof}:
\begin{equation}
    b_{ij} = \sum_{n=1}^{10} G_n(\mu_1, \dots, \mu_5, f_k)\, T_{ij}^{(n)}(\hat{\mathbf{S}}, \hat{\boldsymbol{\Omega}}),
    \label{eq:subfilter_10tensor}
\end{equation}
where $\hat{\mathbf{S}}$ and $\hat{\boldsymbol{\Omega}}$ are evaluated from the resolved velocity field $\bar{\mathbf{u}} = \mathcal{P}\mathbf{u}$. In the RANS limit ($f_k \to 1$), Eq.~\eqref{eq:subfilter_10tensor} recovers the near-equilibrium componential approximation of Eq.~\eqref{eq:rans_10tensor}; as the filter cut-off narrows into the Kolmogorov inertial range ($f_k \to 0$), the neural scalar functions $G_n$ collapse smoothly onto universal inertial-range constants ($G_1 \approx -0.09$, $G_{n>1} \to 0$), guaranteeing seamless physical consistency across the entire RANS-to-DNS bridging spectrum \cite{girimaji2006pans, taghizadeh2024pof}.

\subsection{Seen Flow Categories: Non-Intrusive Reduced-Order Modeling and Implementation Recipe}

When engineering design or optimization is restricted to a \textbf{known flow category} $\mathcal{T}$---a family of parameterized geometries sharing identical topological features---the spatial memory kernel $K_{ij}(\mathbf{x}; \boldsymbol{\Pi})$ remains topodynamically invariant and varies smoothly across a low-dimensional manifold governed by Buckingham $\boldsymbol{\Pi}$ groups \cite{girimaji2026nsr}. Under these conditions, \emph{Pathway~(ii)} bypasses step-by-step causal PDE integration altogether and memorizes the stationary solution manifold directly via Non-Intrusive Reduced-Order Modeling (NIROM) \cite{hesthaven2018, taira2017}.

Beyond reconstructing time-averaged fields, in-class affine flow families also permit non-intrusive synthesis of \textbf{unsteady time evolution at a statistically stationary state} \cite{girimaji2026nsr}. Because the normalized reference memory kernel $\boldsymbol{\mathcal{K}}^*(\boldsymbol{\xi}, \boldsymbol{\xi}', \tau^*)$ preserves its spatio-temporal propagator structure across $\boldsymbol{\Pi}$, pooling \emph{multiple instantaneous realizations} from all training experiments onto the canonical reference domain $\hat{\Omega}$ yields a shared set of \textbf{collective spatial POD modes} $\{\phi_{i,n}(\boldsymbol{\xi})\}$ (alongside stress modes $\{\psi_{ij,m}(\boldsymbol{\xi})\}$). On this collective uncentered eigenbasis \cite{lumley1967, sirovich1987, girimaji2026nsr}, both the statistically stationary moments and the instantaneous velocity evolution $u_i^*(\boldsymbol{\xi}, t^*; \boldsymbol{\Pi})$ expand as:
\begin{equation}
    \bar{u}_i^*(\boldsymbol{\xi}; \boldsymbol{\Pi}) \approx \sum_{n=1}^{N_U} \bar{a}_n(\boldsymbol{\Pi})\, \phi_{i,n}(\boldsymbol{\xi}), \qquad
    \tau_{ij}^*(\boldsymbol{\xi}; \boldsymbol{\Pi}) \approx \sum_{m=1}^{N_R} b_m(\boldsymbol{\Pi})\, \psi_{ij,m}(\boldsymbol{\xi}), \qquad
    u_i^*(\boldsymbol{\xi}, t^*; \boldsymbol{\Pi}) \approx \sum_{n=1}^{N_u} a_n(t^*; \boldsymbol{\Pi})\, \phi_{i,n}(\boldsymbol{\xi}),
    \label{eq:nirom_pod}
\end{equation}
where $t^* \equiv t\, U_{\text{ref}} / L_{\text{ref}}$ denotes dimensionless time. In the collective multi-realization expansion, the leading mode ($n=1$) captures the time-averaged base flow ($\bar{a}_1(\boldsymbol{\Pi})$), while the top coherent mode pairs ($2 \le n \le N_u$) represent traveling or oscillating coherent structures (such as vortex shedding or jet flapping). Because characteristic oscillation frequencies shift across operating conditions, the temporal modal coefficients $a_n(t^*; \boldsymbol{\Pi})$ are parameterized at statistical stationarity via configuration-dependent dimensionless phases $\varphi_n(t^*; \boldsymbol{\Pi}) = 2\pi St_n(\boldsymbol{\Pi})\, t^* + \theta_n(\boldsymbol{\Pi})$ rather than unscaled physical time:
\begin{equation}
    a_n(t^*; \boldsymbol{\Pi}) = \bar{a}_n(\boldsymbol{\Pi}) + A_n(\boldsymbol{\Pi}) \cos\!\big(2\pi St_n(\boldsymbol{\Pi})\, t^* + \theta_n(\boldsymbol{\Pi})\big) + \eta_n(t^*; \boldsymbol{\Pi}),
    \label{eq:nirom_time_coeff}
\end{equation}
where $\bar{a}_n(\boldsymbol{\Pi})$ is the mean offset, $A_n(\boldsymbol{\Pi})$ is the coherent modal amplitude, $St_n(\boldsymbol{\Pi}) = f_n L_{\text{ref}} / U_{\text{ref}}$ is the modal Strouhal number, $\theta_n(\boldsymbol{\Pi})$ is the inter-modal phase angle locking quadrature mode pairs, and $\eta_n(t^*; \boldsymbol{\Pi})$ denotes optional low-dimensional stochastic background fluctuations calibrated to preserve modal variance. Evaluating Eqs.~\eqref{eq:nirom_pod}--\eqref{eq:nirom_time_coeff} enables real-time, zero-solver unsteady ``animation'' of coherent flow dynamics across $\boldsymbol{\Pi}$.

Constructing an interpretable, solver-independent NIROM surrogate for both stationary statistics and unsteady time evolution within a certified Flow Category $\mathcal{T}$ follows a concise five-step recipe:
\begin{enumerate}
    \item \textbf{Affine Reference Mapping \& Multi-Realization Normalization:} Map multiple time-resolved realizations from each training configuration $k \in \{1, \dots, N_e\}$ onto the canonical dimensionless reference domain $\boldsymbol{\xi} = \mathbf{A}(\boldsymbol{\Pi}_k)\mathbf{x} + \mathbf{b}$ and normalize instantaneous velocities, stationary stresses, and time by characteristic scales ($u_{i,k}^* = u_i / U_{\text{ref},k}$, $\tau_{ij,k}^* = \tau_{ij} / U_{\text{ref},k}^2$, $t^* = t\, U_{\text{ref},k} / L_{\text{ref},k}$), interpolating onto a shared comparison grid without extrapolation.
    \item \textbf{Uncentered Collective Cross-Parameter POD:} Pool all normalized realizations across all experiments into a single uncentered snapshot ensemble---avoiding cross-experiment mean subtraction, which creates non-physical composite fields that attenuate shifting coherent structures---and solve the collective method-of-snapshots eigenvalue problem to extract invariant spatial modes $\{\phi_{i,n}(\boldsymbol{\xi})\}$ and $\{\psi_{ij,m}(\boldsymbol{\xi})\}$ with deterministic sign orientation, retaining the top energy-containing modes ($n \le N_u$).
    \item \textbf{Buckingham $\boldsymbol{\Pi}$ Reduction \& Modal Parameter Extraction:} Identify the minimal set of independent dimensionless governing groups $\boldsymbol{\Pi} = (\Pi_1, \dots, \Pi_p)$ via dimensional analysis and extract per-experiment stationary coefficients $\{\bar{a}_n, b_m\}$ alongside spectral/Hilbert-transform trajectory parameters $\{A_n, St_n, \theta_n\}$ for the selected unsteady modes.
    \item \textbf{Collinearity-Controlled Parameter Regression \& Validation:} Learn closed-form algebraic mappings $\boldsymbol{\Pi} \mapsto \{\bar{a}_n(\boldsymbol{\Pi}), b_m(\boldsymbol{\Pi}), A_n(\boldsymbol{\Pi}), St_n(\boldsymbol{\Pi}), \theta_n(\boldsymbol{\Pi})\}$ (and kinematic--stress modal couplings $b_m = g_m(\bar{a}_1, \dots, \bar{a}_{N_U})$) via forward stepwise regression constrained by variance-inflation/condition-number thresholds, verifying parameter stability through Monte Carlo leave-$p$-out cross-validation.
    \item \textbf{Zero-Solver Stationary Synthesis \& Unsteady Animation:} For any target design point $\boldsymbol{\Pi}_{\text{new}}$ within the certified parameter domain, evaluate the regressed modal parameters and reconstruct either the stationary 3D mean/stress fields or the unsteady coherent time evolution $u_i^*(\boldsymbol{\xi}, t^*; \boldsymbol{\Pi}_{\text{new}})$ instantaneously via Eqs.~\eqref{eq:nirom_pod}--\eqref{eq:nirom_time_coeff}, subject to kinematic divergence-free and realizability checks.
\end{enumerate}

Two physical caveats govern this unsteady NIROM extension. First, Eq.~\eqref{eq:nirom_time_coeff} applies specifically to \emph{statistically stationary attractors} (such as post-Hopf periodic limit cycles or stationary turbulent states) within a parameter domain free of further structural bifurcations (Criterion~4 in Section~\ref{sec:flow_category}). Second, because NIROM surrogates fit frame-dependent velocity fields directly from collective memory without enforcing differential conservation laws, both stationary field synthesis and unsteady animation constitute non-intrusive trajectory playback on a known attractor: they exhibit \textbf{zero physical causality} and cannot simulate transient spin-up dynamics or be extrapolated outside their trained topological flow category \cite{girimaji2024njp, girimaji2026nsr}.

\subsection{Mathematical Definition and Criteria for a Valid Flow Category}
\label{sec:flow_category}

To prevent unphysical extrapolation of non-intrusive surrogates, we formulate the exact mathematical boundaries of a valid Flow Category $\mathcal{T}$ directly from the Mori--Zwanzig memory kernel in Eq.~\eqref{eq:mz_kernel} \cite{girimaji2026nsr}. In general non-equilibrium flow families, varying geometric parameters $\boldsymbol{\Pi}$ alters vortex trajectories and curves streamlines non-uniformly, causing the spatial memory kernel to change structure ($\mathbf{K}[\mathcal{F}_1] \neq \mathbf{K}[\mathcal{F}_2]$) \cite{gouasmi2017, girimaji2024njp}. However, when geometric variations across configurations are restricted to anisotropic linear affine transformations, $\boldsymbol{\xi} = \mathbf{A}(\boldsymbol{\Pi})\mathbf{x} + \mathbf{b}$, the coordinate mapping possesses a spatially uniform Jacobian matrix $\mathbf{J} = \mathbf{A}$.

Because $\mathbf{J}$ is constant across space, the metric connection terms (Christoffel symbols $\Gamma_{ij}^k = \nabla_{\mathbf{x}}^2 \xi_k$) vanish identically ($\nabla_{\mathbf{x}}^2 \boldsymbol{\xi} = \mathbf{0}$) \cite{spivak1979}. Spatial differential operators in reference space therefore retain spatially uniform coefficients, avoiding position-dependent curvature artifacts that distort viscous diffusion ($\nu \nabla^2$) or strain-rate alignments \cite{spivak1979, marsden1983}. Letting $\tau \equiv t - s$ denote the elapsed memory lag and $\mathbf{K}$ denote the tensorial form of $K_{ij}$ in Eq.~\eqref{eq:mz_kernel}, memory transport advecting along streamlines scales homogeneously in each coordinate direction, causing the spatial memory kernel to factor cleanly into a parameter-dependent scalar prefactor $C(\boldsymbol{\Pi})$ and a universal, parameter-independent reference kernel $\boldsymbol{\mathcal{K}}^*$ \cite{lucor2004, gouasmi2017}:
\begin{equation}
    \mathbf{K}(\mathbf{x}, \mathbf{x}', \tau; \boldsymbol{\Pi}) = C(\boldsymbol{\Pi}) \, \mathbf{J} \, \boldsymbol{\mathcal{K}}^*(\boldsymbol{\xi}, \boldsymbol{\xi}', \tau^*) \, \mathbf{J}^T.
    \label{eq:affine_kernel}
\end{equation}
This factorization demonstrates that memory transport in self-affine flows is \emph{spatially self-similar}: parameter variations rescale the magnitude and spatial extent of non-local history effects, but leave the structural dynamics of the orthogonal propagator invariant.

Mathematically, a parameterized flow family $\Omega(\boldsymbol{\Pi})$ constitutes a valid Flow Category $\mathcal{T}$ admitting non-intrusive reduced-order parameterization if and only if it satisfies four strict conditions:
\begin{enumerate}
    \item \textbf{Affine Mapping \& Kernel Self-Similarity:} Physical domains $\Omega(\boldsymbol{\Pi})$ must admit a continuous, linear affine mapping $\boldsymbol{\xi} = \mathbf{A}(\boldsymbol{\Pi})\mathbf{x} + \mathbf{b}$ onto a canonical reference domain $\hat{\Omega}$ \cite{spivak1979, marsden1983}. Vanishing second derivatives ($\nabla_{\mathbf{x}}^2 \boldsymbol{\xi} = \mathbf{0}$) ensure zero Christoffel symbols, guaranteeing that the Mori--Zwanzig memory kernel factors into a scalar prefactor $C(\boldsymbol{\Pi})$ and an invariant reference kernel $\boldsymbol{\mathcal{K}}^*(\boldsymbol{\xi}, \boldsymbol{\xi}', \tau^*)$ as in Eq.~\eqref{eq:affine_kernel} \cite{lucor2004, gouasmi2017}.

    \item \textbf{POD Eigenbasis Invariance:} For statistically stationary flows, the spatial auto-correlation operator of the time-integrated normalized reference kernel $\bar{\boldsymbol{\mathcal{K}}}^*(\boldsymbol{\xi}, \boldsymbol{\xi}') \equiv \int_0^\infty \boldsymbol{\mathcal{K}}^*(\boldsymbol{\xi}, \boldsymbol{\xi}', \tau^*)\, d\tau^*$,
    \begin{equation}
        \mathbf{R}(\boldsymbol{\xi}, \boldsymbol{\xi}') = \int_{\hat{\Omega}} \bar{\boldsymbol{\mathcal{K}}}^*(\boldsymbol{\xi}, \boldsymbol{\eta})\, \bar{\boldsymbol{\mathcal{K}}}^{*T}(\boldsymbol{\eta}, \boldsymbol{\xi}')\, d\boldsymbol{\eta},
        \label{eq:pod_autocorr}
    \end{equation}
    is invariant across $\boldsymbol{\Pi}$, ensuring that the uncentered POD spatial eigenmodes $\{\phi_{i,n}(\boldsymbol{\xi})\}$ and $\{\psi_{ij,m}(\boldsymbol{\xi})\}$ remain fixed in reference coordinates while parameter variations enter exclusively through smooth modal amplitudes and temporal phase trajectories \cite{hesthaven2018, girimaji2026nsr}.

    \item \textbf{Spectral Rank Stability:} The singular value spectrum of the cross-parameter snapshot ensemble must remain compact and rapidly decaying across $\boldsymbol{\Pi}$, maintaining a low effective modal rank ($N_U \sim \mathcal{O}(1)$--$\mathcal{O}(10)$, e.g., $N_U \le 6$ in canonical jet-impingement families) without high-wavenumber spectral broadening \cite{taira2017, girimaji2026nsr}.

    \item \textbf{Absence of Structural Bifurcations:} The governing Navier--Stokes solution branch must remain topologically smooth across the parameter domain $\boldsymbol{\Pi}$, exhibiting no pitchfork, Hopf, or shedding-mode bifurcations that induce discontinuous jumps in coherent structure topology or memory kernel symmetry \cite{lucor2004, girimaji2026nsr}.
\end{enumerate}

\section{Operational Decision Roadmap and Conclusions}
\label{sec:roadmap}

\begin{figure}[htbp]
\centering
\begin{tikzpicture}[
    every node/.style={font=\footnotesize},
    startbox/.style={rectangle, rounded corners=4pt, draw=black!80, fill=gray!12, thick, align=center, text width=6.4cm, inner sep=6pt},
    decision/.style={diamond, aspect=2.35, draw=blue!70!black, fill=blue!5, thick, align=center, text width=2.85cm, inner sep=1.5pt},
    action_eq/.style={rectangle, rounded corners=4pt, draw=teal!80!black, fill=teal!7, thick, align=left, text width=4.55cm, minimum height=4.05cm, inner sep=6pt},
    action_causal/.style={rectangle, rounded corners=4pt, draw=orange!80!black, fill=orange!8, thick, align=left, text width=4.55cm, minimum height=4.05cm, inner sep=6pt},
    action_nirom/.style={rectangle, rounded corners=4pt, draw=purple!75!black, fill=purple!7, thick, align=left, text width=4.55cm, minimum height=4.05cm, inner sep=6pt},
    arrow/.style={-{Stealth[length=2.4mm, width=1.7mm]}, thick, draw=black!80}
]

\node[startbox] (start) at (0, 0) {\textbf{\small Turbulence Modeling Task}\\Evaluate Flow State ($De_{\text{turb}} \equiv Sk/\epsilon$) \& User Intent};

\node[decision] (dec_eq) at (0, -1.85) {Near Equilibrium?\\($De_{\text{turb}} \ll 1$)};

\node[decision] (dec_cat) at (2.65, -3.70) {Known Category $\mathcal{T}$?\\(Regimes III \& IV)};

\node[decision] (dec_crit) at (5.35, -5.50) {Valid Category $\mathcal{T}$?\\(4 Criteria Met)};

\node[action_eq] (box_eq) at (-5.25, -9.05) {
    \centering\textbf{\small Intrusive Local Closures}\\
    {\scriptsize (RANS / 10-Tensor Integrity Basis)}\\[4pt]
    \raggedright\scriptsize
    $\bullet$ $\tau_{ij} \approx \tau_{ij}^L$ ($\tau_{ij}^{NL} \to 0$)\\
    $\bullet$ 10-tensor componential approx. for $b_{ij}$ (omits dimensionality)\\
    $\bullet$ RDT-bounded training data\\
    $\bullet$ \textbf{Constraints:} Kinematic + Dynamic
};

\node[action_causal] (box_causal) at (0, -9.05) {
    \centering\textbf{\small Pathway (i): Causal Scale Resolution}\\
    {\scriptsize (PANS / LES + Rules-Augmented AI)}\\[4pt]
    \raggedright\scriptsize
    $\bullet$ Unseen topology ($\mathcal{F}_{\text{unseen}}$); $\tau_{ij}^{NL}$ unknown\\
    $\bullet$ Resolves memory \& eddy dimensionality; $f_k \to 0$\\
    $\bullet$ Same 10-tensor basis closes \emph{subgrid} $b_{ij}$\\
    $\bullet$ \textbf{Constraints:} Intrusive (Kin. + Dyn.)
};

\node[action_nirom] (box_nirom) at (5.25, -9.05) {
    \centering\textbf{\small Pathway (ii): Topological Memorization}\\
    {\scriptsize (NIROM + Collective Uncentered POD)}\\[4pt]
    \raggedright\scriptsize
    $\bullet$ Self-similar kernel $\boldsymbol{\mathcal{K}}^*$ ($\nabla^2\boldsymbol{\xi}=\mathbf{0}$)\\
    $\bullet$ Stationary synthesis \& unsteady animation via $a_n(t^*;\boldsymbol{\Pi})$\\
    $\bullet$ Zero physical causality (no extrapolation)\\
    $\bullet$ \textbf{Constraints:} Non-Intrusive (Kin. Only)
};

\draw[arrow] (start.south) -- (dec_eq.north);

\draw[arrow] (dec_eq.west) -| (box_eq.north)
    node[pos=0.25, above, font=\scriptsize] {\textbf{Yes} ($\tau_{ij}^{NL} \to 0$)};

\draw[arrow] (dec_eq.east) -| (dec_cat.north)
    node[pos=0.30, above, font=\scriptsize] {\textbf{No} ($De_{\text{turb}} \gg 1$)};

\draw[arrow] (dec_cat.west) -| ($(box_causal.north)+(-0.75,0)$)
    node[pos=0.32, above, font=\scriptsize] {\textbf{No} ($\mathcal{F}_{\text{unseen}}$)};

\draw[arrow] (dec_cat.east) -| (dec_crit.north)
    node[pos=0.32, above, font=\scriptsize] {\textbf{Yes} ($\Omega(\boldsymbol{\Pi})$)};

\draw[arrow] (dec_crit.south) -- ($(box_nirom.north)+(0.10,0)$)
    node[pos=0.45, right, font=\scriptsize] {\textbf{Yes}};

\draw[arrow] (dec_crit.west) -| ($(box_causal.north)+(0.75,0)$)
    node[pos=0.28, above, font=\scriptsize] {\textbf{No} (Violated)};

\end{tikzpicture}
\caption{Operational decision roadmap mapping turbulent flow regimes ($De_{\text{turb}}$) and computational intent onto intrusive near-equilibrium closures, causal scale-resolving simulations, and non-intrusive reduced-order models.}
\label{fig:roadmap}
\end{figure}
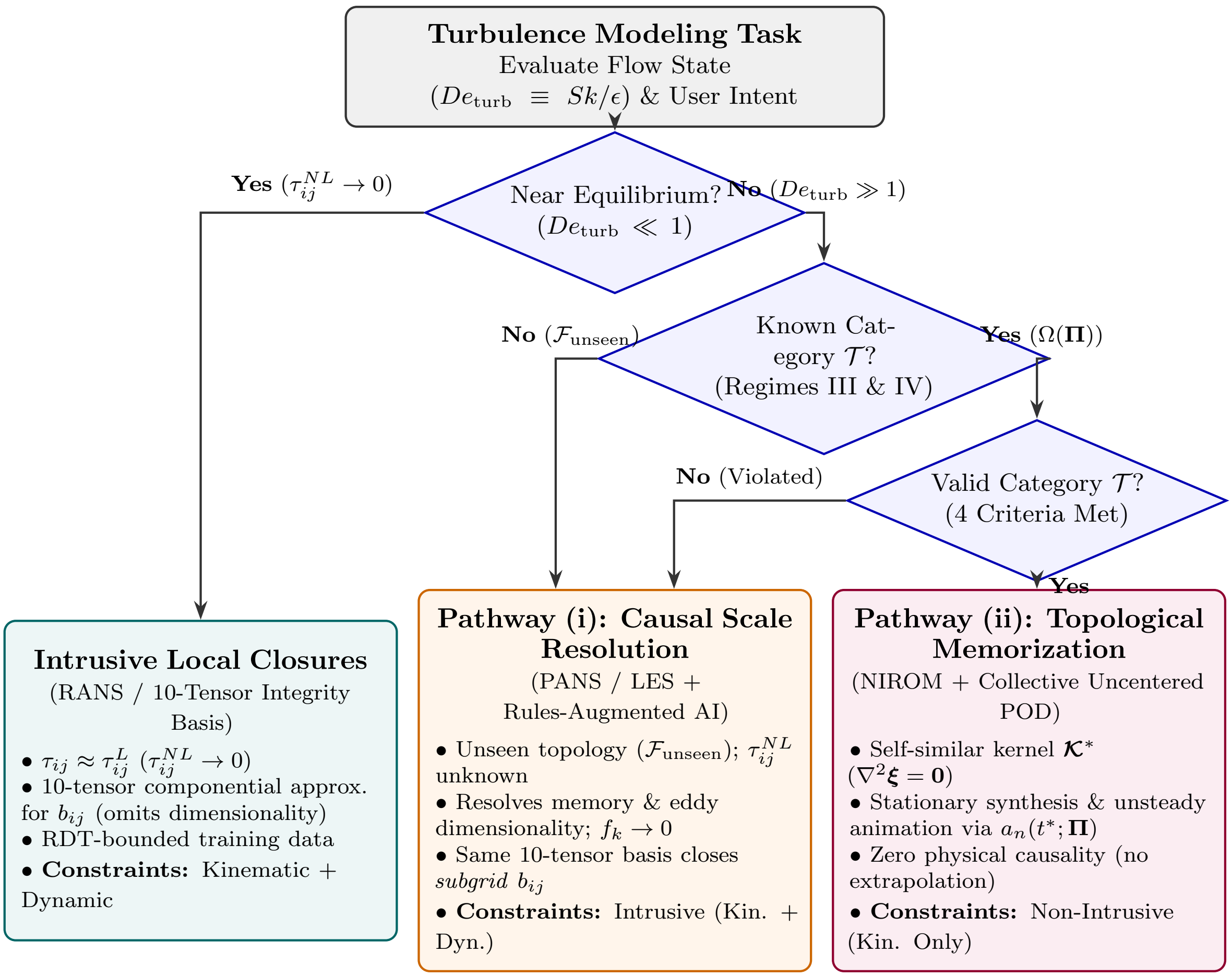

Synthesizing the Mori--Zwanzig transport decomposition with computational intent yields a unified operational decision roadmap (illustrated in Figure~\ref{fig:roadmap} and summarized across turbulence regimes in Table~\ref{tab:unified_taxonomy}). Rather than treating machine learning as a monolithic replacement for physical closures, practitioners must select the modeling architecture dictated by two sequential diagnostic questions: (1) whether the turbulent state resides near statistical equilibrium ($De_{\text{turb}} \ll 1$) or far from equilibrium ($De_{\text{turb}} \gg 1$), and (2) in non-equilibrium regimes, whether the target application involves predictive extrapolation to an unseen geometry ($\mathcal{F}_{\text{unseen}}$) or rapid design interpolation within a certified Flow Category $\mathcal{T}$.

By anchoring AI deployment in the Mori--Zwanzig projection formalism and turbulence structure physics, this framework resolves the apparent conflict between physical rigor and data-driven acceleration. In near-equilibrium flows ($De_{\text{turb}} \ll 1$), kinetic scale separation, Chapman--Enskog asymptotics, and Onsager reciprocity enable generalizable intrusive closures via Pope's ten-tensor integrity basis for the total Reynolds stress---recognized clearly as an acceptable componential engineering approximation that omits single-point eddy dimensionality induced by the elliptic pressure Poisson equation. In non-equilibrium flows ($De_{\text{turb}} \gg 1$), distinguishing predictive extrapolation from within-class interpolation prevents unphysical surrogate deployment: unseen geometries demand first-principles causal scale resolution arbitraged by Rules-Augmented AI---which explicitly captures macroscopic memory and eddy dimensionality while re-enabling the same ten-tensor expansion at the subfilter level---whereas parameterized flow families satisfying the four mathematical criteria of a valid Flow Category admit instantaneous, zero-solver stationary synthesis and unsteady attractor animation via collective, uncentered POD surrogates.

\begin{table}[htbp]
\centering
\caption{Unified taxonomy of data-driven turbulence modeling paradigms across equilibrium and non-equilibrium regimes.}
\label{tab:unified_taxonomy}
\footnotesize
\renewcommand{\arraystretch}{1.35}
\setlength{\tabcolsep}{5pt}
\begin{tabularx}{\textwidth}{>{\raggedright\arraybackslash}p{0.21\textwidth} >{\raggedright\arraybackslash}p{0.22\textwidth} >{\raggedright\arraybackslash}X >{\raggedright\arraybackslash}X}
\toprule
\textbf{Flow Regime \& Timescale} & \textbf{Dominant MZ Physics} & \textbf{User Intent \& Optimal AI Paradigm} & \textbf{Causality \& Required Constraints} \\
\midrule
\textbf{Regimes I \& II:} Near-Equilibrium ($De_{\text{turb}} \ll 1$, $\tau_N \ll \tau_L$) & Local constitutive response ($\tau_{ij} \approx \tau_{ij}^L$; $\tau_{ij}^{NL} \to 0$) & \textbf{Predictive closures across geometries:} Intrusive RANS with 10-tensor componential approx. for \emph{total} $b_{ij}$ (omits dimensionality), trained within RDT bounds. & Statistical correlation absorbs local causation up to single-point pressure-dimensionality floor. \textbf{Intrusive:} Kinematic + dynamic constraints. \\
\midrule
\textbf{Regimes III \& IV:} Unseen Flows ($\mathcal{F}_{\text{unseen}}$) ($De_{\text{turb}} \gg 1$, $\tau_N \gg \tau_L$) & Spatio-temporally non-local memory ($\tau_{ij}^{NL}$ dominates; $K_{ij}$ unknown) & \textbf{Predictive extrapolation to novel topologies:} Causal scale resolution (PANS/LES) with RAAI modulating $f_k$ (captures memory \& dimensionality); 10-tensor basis closes \emph{subfilter} $b_{ij}$. & Causal PDE evolution mandatory; local correlation invalid for total stress. \textbf{Intrusive:} Kinematic + dynamic subgrid consistency. \\
\midrule
\textbf{Regimes III \& IV:} Seen Category ($\mathcal{T}$) ($De_{\text{turb}} \gg 1$, $\tau_N \gg \tau_L$) & Self-similar memory kernel ($\boldsymbol{\mathcal{K}}^*$ invariant under affine $\boldsymbol{\xi}$) & \textbf{Fast design and unsteady animation within a certified Flow Category $\mathcal{T}$:} NIROM with collective uncentered POD ($\boldsymbol{\Pi} \mapsto \{\bar{a}_n, b_m, A_n, St_n, \theta_n\}$). & Purely correlational memorization; zero physical causality. \textbf{Non-Intrusive:} Kinematic only ($\nabla \cdot \bar{\mathbf{u}} = 0$, realizability). \\
\bottomrule
\end{tabularx}
\end{table}



\begin{thebibliography}{36}

\bibitem{girimaji2024njp}
S.~S. Girimaji, ``Turbulence closure modeling with machine learning: a foundational physics perspective,'' \emph{New Journal of Physics}, vol.~26, no.~7, 071201, 2024.

\bibitem{girimaji2026nsr}
S.~S. Girimaji, ``AI meets turbulence modeling: Complex systems perspective,'' \emph{National Science Review}, nwag654, 2026.

\bibitem{kassinos1994}
S.~C. Kassinos and W.~C. Reynolds, ``A structure-based model for the rapid distortion of homogeneous turbulence,'' Tech. Rep. TF-61, Thermosciences Division, Department of Mechanical Engineering, Stanford University, 1994.

\bibitem{kassinos2001}
S.~C. Kassinos, W.~C. Reynolds, and M.~M. Rogers, ``One-point turbulence structure tensors,'' \emph{Journal of Fluid Mechanics}, vol.~428, pp.~213--248, 2001.

\bibitem{girimaji2026neq}
S.~S. Girimaji, ``Coherence as thermodynamic organization: Toward a non-equilibrium turbulence theory,'' \emph{arXiv preprint} arXiv:2607.11817 [physics.flu-dyn], 2026.

\bibitem{monin1971}
A.~S. Monin and A.~M. Yaglom, \emph{Statistical Fluid Mechanics: Mechanics of Turbulence}, Vol.~1. Cambridge, MA: MIT Press, 1971.

\bibitem{hesthaven2018}
J.~S. Hesthaven and S.~Ubbiali, ``Non-intrusive reduced order modeling of nonlinear problems using neural networks,'' \emph{Journal of Computational Physics}, vol.~363, pp.~55--78, 2018.

\bibitem{zwanzig1961}
R.~Zwanzig, ``Memory effects in irreversible thermodynamics,'' \emph{Physical Review}, vol.~124, no.~4, pp.~983--992, 1961.

\bibitem{zwanzig2001}
R.~Zwanzig, \emph{Nonequilibrium Statistical Mechanics}. New York: Oxford University Press, 2001.

\bibitem{chorin2002}
A.~J. Chorin, O.~H. Hald, and R.~Kupferman, ``Optimal prediction with memory,'' \emph{Physica D: Nonlinear Phenomena}, vol.~166, no.~3--4, pp.~239--257, 2002.

\bibitem{parish2017}
E.~J. Parish and K.~Duraisamy, ``Non-Markovian closure models for large eddy simulations using the Mori--Zwanzig formalism,'' \emph{Physical Review Fluids}, vol.~2, no.~1, 014604, 2017.

\bibitem{pope2000}
S.~B. Pope, \emph{Turbulent Flows}. Cambridge: Cambridge University Press, 2000.

\bibitem{bgk1954}
P.~L. Bhatnagar, E.~P. Gross, and M.~Krook, ``A model for collision processes in gases. I. Small amplitude processes in charged and neutral one-component systems,'' \emph{Physical Review}, vol.~94, no.~3, pp.~511--525, 1954.

\bibitem{rodi1976}
W.~Rodi, ``A new algebraic relation for calculating the Reynolds stresses,'' \emph{ZAMM -- Journal of Applied Mathematics and Mechanics}, vol.~56, no.~3, pp.~T219--T221, 1976.

\bibitem{girimaji1996}
S.~S. Girimaji, ``Fully explicit and self-consistent algebraic Reynolds stress model,'' \emph{Theoretical and Computational Fluid Dynamics}, vol.~8, no.~6, pp.~387--402, 1996.

\bibitem{chapman1970}
S.~Chapman and T.~G. Cowling, \emph{The Mathematical Theory of Non-Uniform Gases}, 3rd ed. Cambridge: Cambridge University Press, 1970.

\bibitem{onsager1931}
L.~Onsager, ``Reciprocal relations in irreversible processes. I.,'' \emph{Physical Review}, vol.~37, no.~4, pp.~405--426, 1931.

\bibitem{gouasmi2017}
A.~Gouasmi, E.~J. Parish, and K.~Duraisamy, ``A priori estimation of memory effects in reduced-order models of nonlinear systems using the Mori--Zwanzig formalism,'' \emph{Proceedings of the Royal Society A}, vol.~473, no.~2205, 20170385, 2017.

\bibitem{taghizadeh2021pof}
S.~Taghizadeh, F.~D. Witherden, Y.~A. Hassan, and S.~S. Girimaji, ``Turbulence closure modeling with data-driven techniques: Investigation of generalizable deep neural networks,'' \emph{Physics of Fluids}, vol.~33, no.~11, 115132, 2021.

\bibitem{batchelor1954}
G.~K. Batchelor and I.~Proudman, ``The effect of rapid distortion of a fluid in turbulent motion,'' \emph{The Quarterly Journal of Mechanics and Applied Mathematics}, vol.~7, no.~1, pp.~83--103, 1954.

\bibitem{townsend1976}
A.~A. Townsend, \emph{The Structure of Turbulent Shear Flow}, 2nd ed. Cambridge: Cambridge University Press, 1976.

\bibitem{kassinos2001}
S.~C. Kassinos, W.~C. Reynolds, and M.~M. Rogers, ``One-point turbulence structure tensors,'' \emph{Journal of Fluid Mechanics}, vol.~428, pp.~213--248, 2001.

\bibitem{pope1975}
S.~B. Pope, ``A more general effective-viscosity hypothesis,'' \emph{Journal of Fluid Mechanics}, vol.~72, no.~2, pp.~331--340, 1975.

\bibitem{mishra2010}
A.~A. Mishra and S.~S. Girimaji, ``Pressure--strain correlation modeling: towards achieving consistency with rapid distortion theory,'' \emph{Physical Review E}, vol.~81, no.~5, 056301, 2010.

\bibitem{mishra2013jfm}
A.~A. Mishra and S.~S. Girimaji, ``Intercomponent energy transfer in incompressible homogeneous turbulence: multi-point physics and amenability to one-point closures,'' \emph{Journal of Fluid Mechanics}, vol.~731, pp.~639--681, 2013.

\bibitem{mishra2017jfm}
A.~A. Mishra and S.~S. Girimaji, ``Toward approximating non-local dynamics in single-point pressure--strain correlation closures,'' \emph{Journal of Fluid Mechanics}, vol.~811, pp.~168--188, 2017.

\bibitem{taghizadeh2020njp}
S.~Taghizadeh, F.~D. Witherden, and S.~S. Girimaji, ``Turbulence closure modeling with data-driven techniques: physical compatibility and consistency considerations,'' \emph{New Journal of Physics}, vol.~22, no.~9, 093023, 2020.

\bibitem{speziale1987}
C.~G. Speziale, ``On nonlinear $K$--$l$ and $K$--$\epsilon$ models of turbulence,'' \emph{Journal of Fluid Mechanics}, vol.~178, pp.~459--475, 1987.

\bibitem{lumley1978}
J.~L. Lumley, ``Computational modeling of turbulent flows,'' \emph{Advances in Applied Mechanics}, vol.~18, pp.~123--176, 1978.

\bibitem{girimaji2004jfm}
S.~S. Girimaji, ``Lower-dimensional manifold modeling of turbulent flows: Realizability and fixed-point stability,'' \emph{Journal of Fluid Mechanics}, vol.~512, pp.~191--210, 2004.

\bibitem{sagaut2006}
P.~Sagaut, \emph{Large Eddy Simulation for Incompressible Flows: An Introduction}, 3rd ed. Berlin: Springer, 2006.

\bibitem{girimaji2006pans}
S.~S. Girimaji, ``Partially-averaged Navier--Stokes method for turbulence: A Reynolds-averaged Navier--Stokes to direct numerical simulation bridging method,'' \emph{Journal of Applied Mechanics}, vol.~73, no.~3, pp.~422--429, 2006.

\bibitem{taghizadeh2024pof}
S.~Taghizadeh, F.~D. Witherden, and S.~S. Girimaji, ``Scale-resolving simulations of turbulent flows with coherent structures: Toward cut-off dependent data-driven closure modeling,'' \emph{Physics of Fluids}, vol.~36, no.~6, 065143, 2024.

\bibitem{taira2017}
K.~Taira, S.~L. Brunton, S.~T.~M. Dawson, C.~W. Rowley, T.~Colonius, B.~J. McKeon, O.~T. Schmidt, S.~Gordeyev, V.~Theofilis, and L.~S. Ukeiley, ``Modal analysis of fluid flows: An overview,'' \emph{AIAA Journal}, vol.~55, no.~12, pp.~4013--4041, 2017.

\bibitem{lumley1967}
J.~L. Lumley, ``The structure of inhomogeneous turbulent flows,'' in \emph{Atmospheric Turbulence and Radio Wave Propagation}, A.~M. Yaglom and V.~I. Tatarski, Eds. Moscow: Nauka, 1967, pp.~166--178.

\bibitem{sirovich1987}
L.~Sirovich, ``Turbulence and the dynamics of coherent structures. Part I: Coherent structures,'' \emph{Quarterly of Applied Mathematics}, vol.~45, no.~3, pp.~561--571, 1987.

\bibitem{spivak1979}
M.~Spivak, \emph{A Comprehensive Introduction to Differential Geometry}, Vol.~2, 2nd ed. Berkeley, CA: Publish or Perish, 1979.

\bibitem{marsden1983}
J.~E. Marsden and T.~J.~R. Hughes, \emph{Mathematical Foundations of Elasticity}. Englewood Cliffs, NJ: Prentice-Hall, 1983.

\bibitem{lucor2004}
D.~Lucor and G.~E. Karniadakis, ``Noisy inflows cause a shedding-mode switching in flow past an oscillating cylinder,'' \emph{Physical Review Letters}, vol.~92, no.~15, 154501, 2004.

\end{thebibliography}
\end{document}